# Bacterial swimming and accumulation on endothelial cell surfaces

Xin-Xin Xu, [†a] Yangguang Tian, [†b] Yuhe Pu, [b] Bingchen Che, [c] Hao Luo, [b] Yanan Liu, [b] Yan-Jun Liu*[a] and Guangyin Jing*[b]

a.  Shanghai Xuhui Central Hospital, Zhongshan-Xuhui Hospital, Shanghai Key Laboratory of Medical Epigenetics, Institutes of Biomedical Sciences, Department of Chemistry, Fudan University, Shanghai 200032, China
b.  School of physics, Northwest University, 710127, Xi'an, China
c.  Department of Chemistry, Key Laboratory of Bioorganic Phosphorus Chemistry & Chemical Biology (Ministry of Education), Tsinghua University, 100084, Beijing, China.

jing@nwu.edu.cn; Yanjun_Liu@fudan.edu.cn.

**Abstract**

Flagellar-driven locomotion plays a critical role in bacterial attachment and colonization of surfaces, contributing to the risks of contamination and infection. Tremendous attempts to uncover the underlying principles governing bacterial motility near surfaces have relied on idealized assumptions of surrounding inorganic boundaries. However, in the context of living systems, the role of cells from tissues and organs becomes increasingly critical, particularly in bacterial swimming and adhesion, yet it remains poorly understood. Here, we propose using biological surfaces composed of vascular endothelial cells to experimentally investigate bacterial motion and interaction behaviors. Our results reveal that bacterial trapping observed on inorganic surfaces is counteractively manifested with reduced radii of circular motion on cellular surfaces, while with two distinct modes of bacterial adhesion: tight adhesion and loose adhesion. Interestingly, the presence of living cells enhances bacterial surface enrichment, and imposed flow intensifies this accumulation via bias-swimming effect. These results surprisingly indicate that physical effects remain the dominant factor regulating bacterial motility and accumulation at the single-cell layer level in vitro, bridging the gap between simplified hydrodynamic mechanisms and complex biological surfaces，with relevance to biofilm formation and bacterial contamination.

**Introduction**

Motile bacteria, a common type of microswimmer, navigate fluidic environments by rotating their flagella, often encountering surface boundaries [1]. This motility enhances nutrient acquisition, aids in immune evasion, and facilitates the formation of bacterial communities [2-5]. At the population level, swimming bacteria exhibit highly uneven spatial distributions, with a pronounced tendency to accumulate on surfaces [6-8]. Such surface accumulation plays a pivotal role in promoting bacterial adhesion, biofilm formation, and infection [9-14]. These phenomena are critical to healthcare, industry, and environmental systems, driving extensive research efforts to uncover the underlying principles governing bacterial motility, adhesion, and colonization on a variety of surfaces [15-19].

Motility of bacteria can root its long history in the context of the consequence of infectious diseases and correspondingly global problem of antibiotic resistance, and beyond biological aspects, swimming bacteria in fluidic environments have also long been of extensive interest to the physics and mechanics communities. With locomotion mechanism, it is critical to investigate how the bacteria interact with their surroundings, particularly with the boundary in a confined space [20]. Hydrodynamic interactions, before and just after the bacteria approach to the surface, need to be carefully examined [21-23]. These swimmer-surface interaction often relies on simplifications and idealized assumptions that the surfaces they encounter are frequently modeled as rigid, impenetrable walls with simplified hydrodynamic boundary conditions [24], and motile bacteria are typically treated as active force and stress generators in fluid with and or without flows [25-28]. Building on these simplifications, seminal studies have uncovered a range of classical motility phenomena on such surfaces. These include circular motion of bacteria with prolonged dwell times on solid surfaces [29], periodic wiggling driven by torque balance under viscous drag [30], self-alignment with boundaries due to hydrodynamic image effects, upstream swimming under shear flow [31], shear trapping caused by flow gradients [32], and drift reorientation induced by the chirality-generated lift force from helical flagella [33].

However, bacterial infections predominantly occur on biological surfaces, such as the membranes of tissue cells, rather than idealized walls [34-36]. These biological surfaces are composed of diverse components, including lipids, proteins, and carbohydrates [37], which confer specific physical properties such as roughness and stiffness [38][39], as well as distinct biological functions [40]. This naturally raises critical questions about the validity of common approximations, such as rigid smooth surfaces, lubrication theories, non-slip conditions, and stress-free assumptions, when applied to living, cell-based membranes. Furthermore, it remains unclear to what extent the simplified laws and classical effects identified on inorganic surfaces apply to biological surfaces, or whether these principles are violated. If deviations do exist, a key challenge is to determine how much of the hydrodynamic behavior observed on idealized surfaces can be extended to, or recovered in, the context of biological surfaces.

To address these questions, we developed living cellular surface as the model system composed of vascular endothelial cells, allowing systematic investigation of bacterial motion and interaction behaviors under shear flow. The results elucidate that cell surfaces reduce the radius of bacterial circular motion by altering the effective height between the bacteria and the surface. Adherent bacteria were observed on cell surfaces and can be categorized into two types. Furthermore, the presence of cells enhances bacterial surface enrichment, primarily through increased adhesion. Flow is imposed manifested the blood stream within capillary vessel, which varies the conditions to further intensify this accumulation due to hydrodynamic interactions between bacteria and shear flow. These accumulations are time-dependent, particularly under flow conditions. Taken together, our results highlight that while biological surfaces modulate bacterial behavior, physical effects remain the predominant factor governing bacterial motility and accumulation at the single-cell layer level in vitro.

**Results and Discussion**

**Circular Motion of Bacteria trapped by Cell surfaces**

The dwell time of bacteria near surfaces play the critical role on bacterial adhesion [41].

It has been found the flagellar bacteria are trapped onto the solid plane with a typical circular motion [42]. The essential assumption has been made that the non-slip boundary condition from the rigid abiotic surface, which generates a net in-plane torque on the bacteria due to the counter-rotating of its body and flagellum [29]. Then, the questions arise whether and how the biological surface such as cell layers change this classic mechanism.

Thus, to mimic the bacterial locomotion on the surfaces of vascular endothelial cells in vitro, we fabricated a PDMS microchannel with dimensions of $100\ \mu m$ in height and $600\ \mu m$ in width, as illustrated in Fig. 1a. The glass bottom of the channel was pre-coated with fibronectin (FN) to promote endothelial cell adhesion. Human umbilical vein endothelial cells (HUVECs) were cultured at varying densities for 6–12 hours to ensure their attachment to the microchannel surface. Fluorescence microscopy was used to image the HUVECs and *Escherichia coli* RP437, as shown in Fig. 1b, which includes merged images of HUVECs, their nuclei, bacteria, and their trajectories over a 3 s. The relationship between cell density and the proportion of cell-covered area is presented in Fig. 1c, revealing a predominantly positive linear correlation.

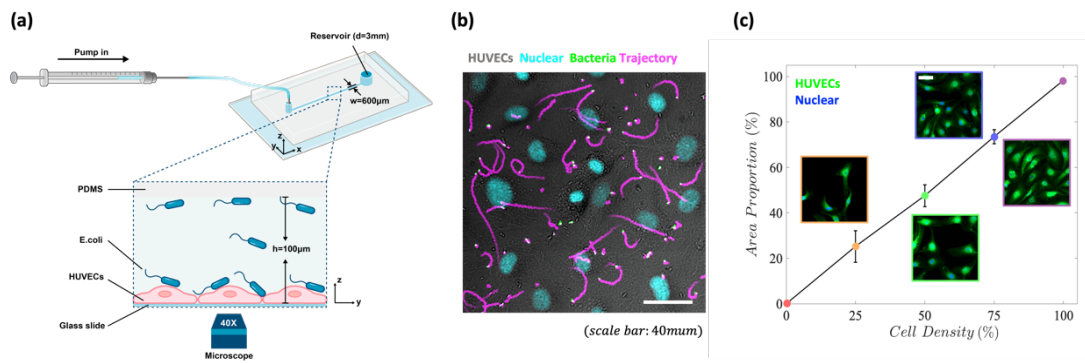

**Figure 1. Experimental Setup for Observing Bacterial Locomotion on Cell Surfaces.**
(a) Schematic representation of the microchannel used in the experiment. The channel has dimensions of $600\ \mu m$ in width and $100\ \mu m$ in height. HUVECs were introduced through the channel entrance and cultured on the bottom surface. *E. coli* RP437 and controlled flow fields were introduced into the channel using a syringe pump. The culture medium was maintained in a reservoir with a $3\ mm$ diameter at the end of the channel. (b) Representative brightfield and fluorescence microscopy images illustrating bacterial trajectories (magenta) over 3 s. Bacteria are shown in green, HUVECs in gray, and nuclei in cyan. Scale bar: $40\ \mu m$. (c) Plot showing the relationship between cell density and the area proportion of HUVECs. Insets depict fluorescence microscopy images of HUVECs (green) and their nuclei (blue) at varying cell densities. Error bars represent the standard deviation of the area proportion calculated from multiple fields of view. Scale bar: 40 μm.

The results show that the circular motion persists on cell surfaces, and this trapped motion is modulated by cell density. Specifically, as the proportion of cell-covered area increases, the radius of the bacterial trajectories decreases, as shown in Fig. 2a. The insets illustrate typical trajectories (orange curves) alongside circles fitted using the average curvature radius (blue dashed lines). Additionally, the angular velocity of the circular motion increases with the cell area proportion, whereas the bacterial linear speeds exhibit a slight decline (Fig. 2b).

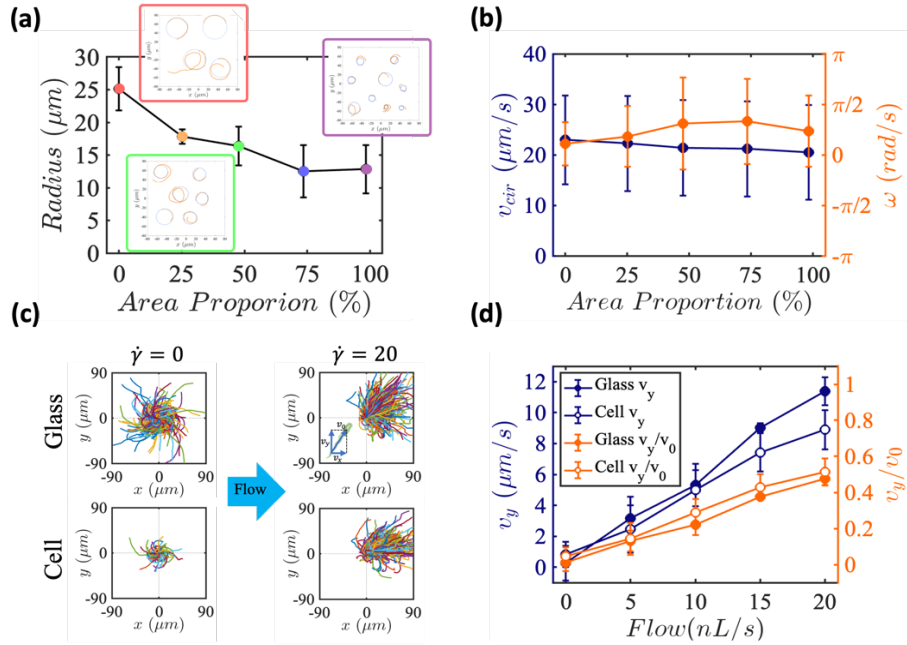

**Figure 2. Regulation and Transformation of Bacterial Circular Motion**
(a) Relationship between the radius of circular motion and cell area proportion. Error bars indicate the standard deviation from approximately 100 trajectories. Insets show representative bacterial circular motion trajectories recorded over 10 s (orange curve) with corresponding fitted trajectories (blue dashed line). (b) Angular velocity (blue) and linear speed (orange) of bacterial circular motion as functions of cell area proportion. (c) Representative bacterial trajectories observed on glass and cell surfaces under shear rates of $\dot{\gamma} = 0$ s$^{-1}$ and $\dot{\gamma} = 20$ s$^{-1}$, recorded over 3 s. (d) Bacterial drift speed ($v_y$, blue) and drift ability ($v_y/v_0$, orange) on cell surfaces influenced by shear flow.

Bacterial circular motion arises from the additional rotational drag forces generated as bacteria swim along a surface, due to the interaction with non-slip boundary conditions imposed by the surface. The counterclockwise (CCW) rotation of the flagella (from tail to head) and the clockwise (CW) torque balancing of the bacterial body generate opposing viscous forces at distinct points of action on the flagella and the body. This interaction generates a net torque, driving bacteria to trace a CW circular trajectory on solid surfaces. [29]. On air-liquid interfaces (complete slip) or more complex polymer interfaces (partial slip), the direction of circular motion reverses from CW to CCW, depending on the slip length [43]. By employing the theory proposed by Lopez and Lauga that the relationship between bacterial circular motion and abiotic interfacial properties [44], then the angular velocity is adapted as,

$$\omega = \frac{3}{256\pi} \frac{q}{\eta h_{eff}^4} \left( \frac{\gamma^2 - 1}{\gamma^2 + 1} + \frac{1 - \lambda}{1 + \lambda} \right).$$

This angular motion depends on several factors, including the bacterial dipole strength (q), the bacterial aspect ratio ($\gamma$), and the slip length of the surface ($\lambda$). By modifying the height in the original theory from ref. [44], here, the effective height of bacteria above the interface ($h_{eff}$) is defined. In our experiments, the bacterial properties, such as aspect ratio ($\gamma$) and dipole strength (q), which corresponds to the linear speed shown in Fig. 2b, remained approximately constant despite changes in cell density. On cell surfaces, the slip length is on the nanoscale [45][46], reasonably treating the surface as a solid for bacterial swimming. Among these parameters, the only variable influenced by cell density is the effective height between the bacteria and the cell surface.

Generally, endothelial surfaces are coated with a glycocalyx, a complex layer composed of proteoglycans, glycoproteins, and glycolipids, with a thickness ranging from tens to hundreds of nanometers [47][48]. Similarly, the height of bacterial circular motion on solid surfaces is approximately ten to hundreds of nanometers [49][50]. Thus, the changes in circular motion can be attribute to the cell surface roughness, which reduces the effective height between the bacteria and the surface. This reduction alters the angular velocity of the circular motion and decreases the radius of the trajectory, in agreement with the results shown in Fig. 2a.

In addition to static fluids, interactions between swimming bacteria and surfaces play a crucial role in bacterial exploration of the surrounding environment within flow conditions. Several rheotactic behaviors, such as upstream swimming and drift motion, have been observed on surfaces, directly contributing to surface accumulation and the potential for infection [51-53]. From a hydrodynamic perspective, Mathijssen et al.

studied the transition of bacterial surface circular motion to drift motion under shear flow. In our experiments, this transition was investigated on complex cell surfaces [54].

As shown in Fig. 2c, bacterial circular trajectories on both glass and cell surfaces completely transition into drift motion under shear flow ($\dot{\gamma} = 20 \text{ s}^{-1}$). However, the trajectories on cell surfaces are shorter due to surface roughness, and the swimming trend in the y-direction is suppressed. Quantitative analysis further supports these findings. As depicted in Fig. 4d, we examined the drift speed ($v_y$) along the y-axis as a function of flow rate. The results indicate that $v_y$ is slightly reduced on cell surfaces. However, when normalized by the bacteria's free-swimming speed ($v_0$) in the absence of flow—representing bacterial drift ability—the drift ability remains constant between cell and glass surfaces. This suggests that the complexities of the cell surface do not significantly affect bacterial drift motion.

The drifting motion is attributed to the lifting force generated by the helical flagella in shear flow. On surfaces, bacteria are reported to adopt a "nose-down" gait with a pitch angle of approximately $10°$ [55]. At a shear rate of $20 \text{ s}^{-1}$, bacteria exhibit downstream direction, with the flow further increasing the pitch angle. Given a bacterial length of $10 \text{ μm}$, this orientation causes the flagella to elevate at least $0.35 \text{ μm}$ above the surface—exceeding the thickness of the glycocalyx [56]. As a result, the roughness of the glycocalyx layer does not interact with the bacterial flagella, reinforcing the robustness of bacterial drift ability even on complex cell surfaces.

In our experiments, vascular endothelial cells were cultured under static conditions, resulting in random orientations without alignment along a specific axis. However, under in vivo conditions, prolonged exposure of endothelial cells to laminar shear stress is essential for vascular homeostasis. This mechanical stimulus activates mechanoreceptors, aligning endothelial cells along the direction of blood flow, regulating intracellular signaling pathways, and inducing the expression of specific genes and proteins [57]. In such anisotropic environments, bacteria are reported to exhibit constrained orbital movements, potentially altering their behavior on biological surfaces [58].

In summary, bacterial swimming on cell surfaces is primarily governed by hydrodynamic interactions. Even on biologically complex surfaces, bacteria exhibit a robust transition from circular to drift motion. At the single-cell layer level in our experiments, the influence of biological factors is relatively weak, with interactions predominantly driven by physical contact. However, as tissues grow and mature, biological functions might become more pronounced. For instance, fully developed vascular endothelial tissue not only acts as a barrier between blood and tissue but also serves as an endocrine organ, regulating blood flow, platelet adhesion, leukocyte activation, and migration [59]. In such cases, the interplay between physical and biological factors remains an intriguing and open question.

**Bacterial Adhesion on Cell Surfaces**

Building on the exploration of bacterial swimming behavior on cell surfaces, we then investigate bacterial adhesion, a crucial aspect of bacterial-host interactions that underpins processes like biofilm formation and infection [60]. In comparison to glass surfaces, we observed a significantly higher number of bacteria adhering to cell surfaces. Based on the movement patterns of their center of mass, these adhered bacteria could be categorized into two types: tightly adherent bacteria, where the center of mass remains almost stationary, and loosely adherent bacteria, where the center of mass moves while the bacteria spin in place (Fig. 3a). We quantify their spatial locations in relation to their movement speeds. Figure 3c shows the average speed of the two types of adherent bacteria as a function of their distance (D) from the cell edge. The results indicate that tightly adherent bacteria primarily cluster along the edges of cells, whereas loosely adherent bacteria are predominantly located on the central regions of the cell surface, away from the edges.

Fig. 3b illustrates the trajectories of the two types of bacteria, plotted separately from the same origin. The displacements of tightly adherent bacteria are consistently shorter than the bacterial body length (2 μm), while the displacements of loosely adherent bacteria are shorter than the total bacterial length, including both the body and flagella (10 μm). We further investigate the behavior of adherent bacteria under flow conditions. Experimentally, after introducing bacteria into the microchannel, a low flow rate (20 nL/s) is applied to remove bacteria in the bulk, leaving behind adherent bacteria and a few free-moving ones. Subsequently, different flow rates (0 − 1000 nL/s) are applied, and the retention ratio of tightly adhered and loosely adhered bacteria (v < 10 μm/s) is quantified. The corresponding shear rates range from near the surface, covering the typical shear conditions encountered in most in vivo vascular environments [61][62].

As shown in Fig. 3d, a low flow rate of 20 nL/s is sufficient to remove approximately 50% of the adherent bacteria, which are predominantly loosely adhered bacteria and a small fraction of free-swimming ones. Assuming that the bacteria experience shear stress over an area of approximately 1 μm², the corresponding shear force is estimated as $F = \mu \cdot \dot{\gamma} \cdot A \approx 0.06$ pN. However, the remaining tightly adhered bacteria are resilient to detachment, even at a flow rate of 1000 nL/s, which corresponds to a shear force of approximately 3 pN. These results suggest that the adhesion force of tightly adhered bacteria exceeds 3 pN, consistent with the range of catch-bond shear stresses associated with type I pili on mannose-coated surfaces, which spans approximately 0 − 10 pN/μm² [63][64].

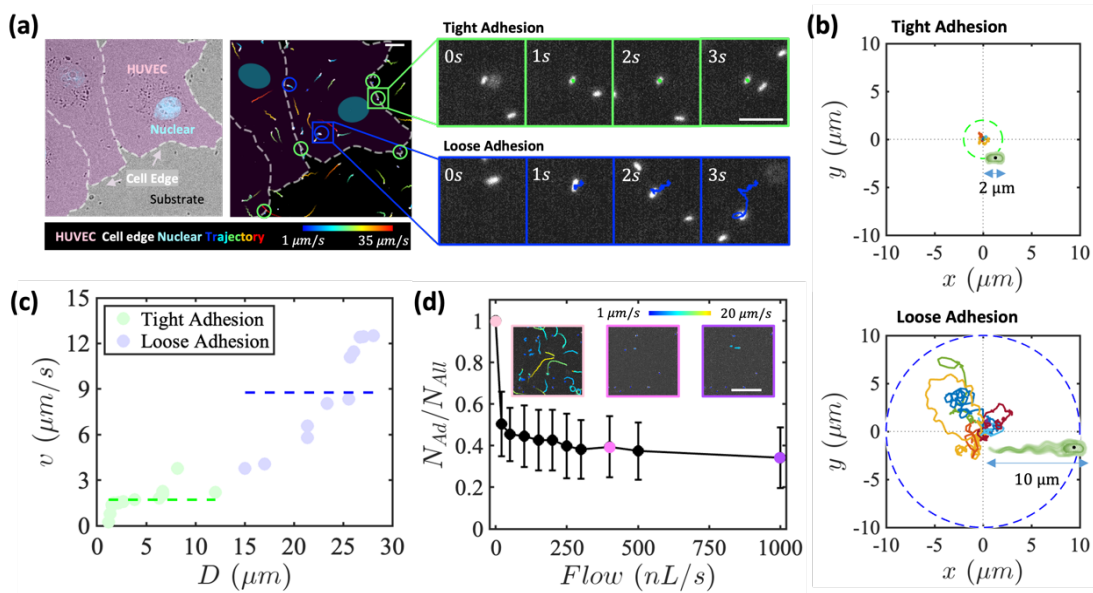

**Figure 3. Bacterial Adhesion on Cell Surfaces**
(a) Representative images showing the positions of nuclei (light blue) and cell boundaries (white) of HUVECs (light pink), overlaid with bacterial trajectories recorded over 0.5 s. Trajectory colors indicate bacterial average speed. Tight adherent bacteria (green circles) are predominantly localized at the cell edge, whereas loose adherent bacteria (blue circles) are distributed far from the edge. The right panel illustrates time-lapse sequences of both adhesion types. Scale bar: 10 μm. (b) Typical trajectories of tight and loose adherent bacteria originating from the same initial point. The dot circles are plotted with the radius of bacterial body and flagella. (c) Average speed ($v$) as a function of distance (D) from the bacterial body to the cell edge. Dashed lines indicate the mean speeds for each adhesion mode. (d) The exist ratio of bacteria eroded by the flow. Insets show trajectories of bacteria remaining on a cell surface with 94% area coverage after flow erosion. Scale bar: 50 μm.

Next, we discuss the possible mechanisms by which bacteria transition from free-swimming to adhesion, a process critical for understanding bacterial colony formation and the onset of infections in living organisms [65][66]. Bacterial adhesion is a complex process governed by physical, chemical, and biological factors [67]. In addition to physical interaction including van der Waals forces, electrostatic interactions, and hydrophobic interactions near surfaces, bacteria with flagella typically initiate adhesion through a weak, flagella-mediated process, which is generally considered reversible. Once bacteria persist on a surface for a certain period, they can sense the presence of the interface through various mechanisms, such as body deformation, flagellar load, or ciliary contraction resistance [68]. These signals stimulate the production of adhesins, facilitating a transition from reversible to irreversible adhesion [69][70]. Adhesins, which exhibit adhesion strengths in the ~ μN range, are commonly found at the tips of pili in gram-negative bacteria or on the cell surface of gram-positive bacteria [71]. When combined with the range of bacterial adhesion forces observed in our flow experiments, these findings suggest that tight adhesion corresponds to body-mediated irreversible adhesion, whereas loose adhesion represents a reversible adhesion process.

Furthermore, many adhesin-mediated adhesion processes are reported to involve catch bonds, a crucial mechanism that enables bacteria to maintain adhesion under flow conditions [72]. For instance, the adhesion of type I pili in pathogenic *E. coli* to mannose on bladder cell surfaces, as well as the specific binding of S. aureus MSCRAMMs to collagen or fibrinogen on endothelial cells, are classic examples [73][74]. In these catch-bond adhesion processes, the lifetime of receptor-ligand bonds increases under optimal shear forces, thereby enhancing adhesion strength. This shear-enhanced adhesion enables bacteria to endure flow conditions while maintaining the flexibility to detach and resume motility in static environments. Such a mechanism allows bacteria to strike a balance between stable colonization and efficient dispersion, facilitating their survival and adaptability in dynamic environments [75][76].

**Bacterial Accumulation Under Flow Erosion**

Bacteria, from physics point of view, due to the dipole structure of the fluid flow generated during their swimming, can become trapped on surfaces, leading to highly uneven spatial distributions. This typically results in bacteria accumulating on surface, significantly increasing the likelihood of biofilm formation and infection [77]. On biological surfaces, as discussed in the previous sections, surface complexities influence bacterial behaviors, including modifications in swimming patterns and the initiation of adhesion. Taking all these factors into account, this section focuses on how surface complexities influence total surface number of bacteria, particularly in relation to its dependence on cell density and temporal dynamics under both flow and no-flow conditions.

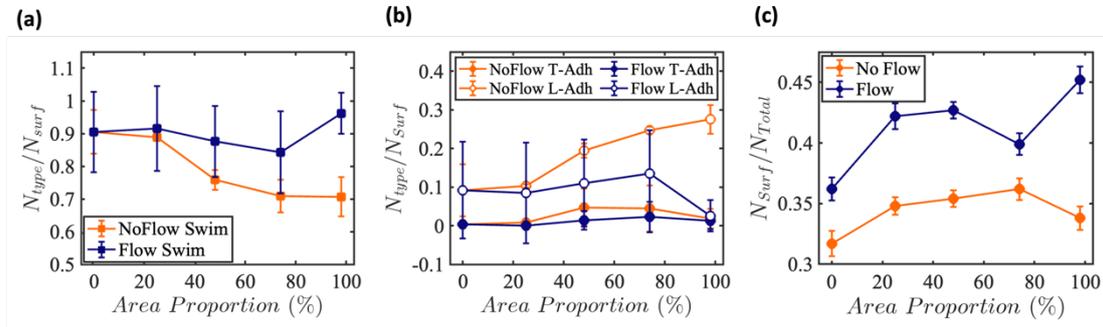

**Figure 4. Surface Accumulation of Bacteria Turned by Cell Density.**
(a) Ratio of swimming bacteria to total surface bacteria as a function of the proportion of the cell surface area in static (orange) and flow (blue) conditions. (b) Ratios of tight adherent (T-Adh) and loose adherent (L-Adh) bacteria to total surface bacteria, plotted against the proportion of the cell surface area. (c) Division of surface bacterial numbers ($N_{surf}$, 0−10 μm) and total bacterial numbers (0−100 μm) as a function of the proportion of the cell surface area under static (orange) and flow (blue) conditions. Flow parameters: 20 nL/s, $\dot{\gamma} = 20$ s$^{-1}$.

Initially, we measure how bacterial surface accumulation varies with cell density under both flow and no-flow conditions. Experimentally, bacteria are introduced into a PDMS channel with a height of 100 μm. Bacterial movement is recorded across z-stacks with a height step of 5 μm. Given that the cell surface height fluctuates within approximately 10 μm, we summed bacterial counts at heights of 0, 5, and 10 μm to represent the surface number ($N_{surf}$), while the total bacterial count across all heights is recorded as We further categorized bacterial counts at heights of 0–10 μm into three types: swimming, tightly adhered, and loosely adhered ($N_{type}$). The proportion of each type relative to the total surface bacteria is calculated as $N_{type}/N_{surf}$, while the normalized surface number was defined as $N_{surf}/N_{Total}$. Both variables are analyzed as functions of the cell area proportion, as shown in Fig. 4. The results indicate that without flow, the number of swimming bacteria decreases while the number of adherent bacteria increases with increasing cell density, leading to a rise in the total bacterial count on the surface. Under flow, although the proportions of swimming and adhered bacteria change only slightly, the total surface bacterial number increases significantly.

Without flow, the increase in surface bacterial numbers is clearly driven by the rise in adherent bacteria. However, under flow conditions, the degree of loose adhesion decreases, consistent with our earlier discussion that loosely adhered bacteria are more easily washed away by the flow. The overall increase in surface bacterial numbers across all cell densities under flow is attributed to the hydrodynamic interactions between bacteria and the shear flow, which can be divided into two key effects: drift motion and shear trapping [78][32].Drift motion along the z-direction is induced by shear stress along the *x*-axis, which enhances the proportion of bacteria approaching the surface. Shear trapping, on the other hand, occurs when shear stress along the z-axis exerts an additional trapping effect on rod-shaped bacteria, particularly in regions of high shear ($0-10$ μm). Together, these effects intensify bacterial surface accumulation under flow erosion.

Next, the time-dependent accumulation of bacteria is examined. Experimentally, we analyze the normalized bacterial count, defined as the bacterial number at time t, N (t), divided by the bacterial number at the initial time, N (0), as a function of time in the same plane ( 0 μm ). In flow conditions, the bacterial count is measured while maintaining a continuous injection of bacterial suspension. As shown in Figure 5, the proportions of swimming and adhered bacteria remain unchanged over time, regardless of whether the system was stationary or flowing. However, for the total bacterial number, time-dependent accumulation is observed in both cases. Without flow, the bacterial number increased slightly over time, whereas under flow, the increase is significantly more pronounced. This enhanced accumulation under flow can be attributed to the shear flow bringing new bacteria to the surface, some of which become trapped, leading to an overall rise in surface bacterial numbers.

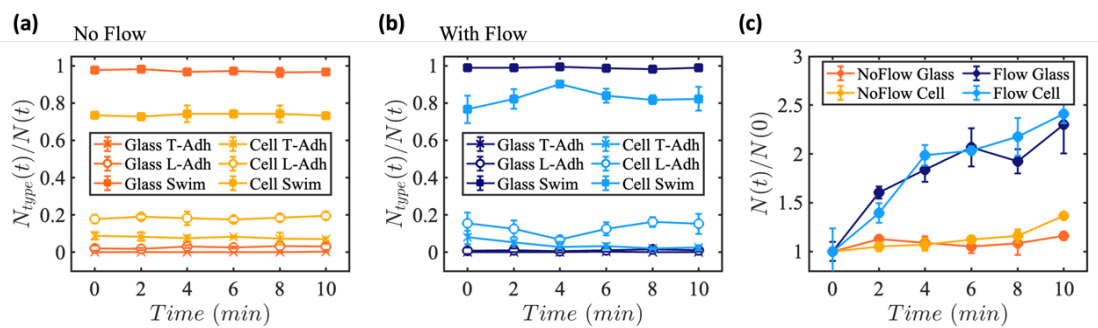

**Figure 5. Bacterial Accumulation Over Time**
(a) Ratios of different bacterial types (swimming, tight adherent, and loose adherent) to the total number of surface bacteria (0 μm), $N_{type}/N_{surf}$, as a function of time under static conditions. (b) Ratios of different bacterial types to the total number of surface bacteria over time under flow conditions (20 nL/s, $\dot{\gamma} = 20$ s$^{-1}$). (c) Normalized bacterial numbers ($N(t)/N(0)$) over time under static (orange) and flow (blue) conditions, showing the influence of flow on bacterial accumulation.

The accumulation of bacteria and other self-propelled microorganisms on surfaces has been a topic of significant interest among researchers since 1963 [6]. The total number of bacteria accumulated on a surface depends on the balance between the trapping and escaping rates. [8][79]. In the absence of external flow, the key factors contributing to the trapping effect include hydrodynamic attraction and the steric effect. The escaping rate primarily results from rotational diffusion caused by bacterial tumbling [80]. Although, the tumbling frequency is inhibited on surface, occurring at approximately 0.5 times per second—half the frequency observed in the bulk fluid [81]. The competition between trapping and escaping leads to bacterial accumulation on surfaces and results in time-dependent enrichment.

Upon introducing flow, bacteria, due to their rod-like shape, exhibit periodic rotation with the classic Jeffery orbits in the shear flow [82][83]. The rotating and stretching characteristics of the shear flow alter the orientation of the bacteria, introducing an additional rotational diffusion that aids in their escape from the surface.

Additionally, shear flow has another effect on rod-shaped particles: shear trapping, where the particles align with the flow direction and become trapped in the same shear plane. For instance, shear stress along the *z*-axis can trap bacteria in the high shear regions near the top and bottom surfaces. Furthermore, due to the helical nature of bacterial flagella, shear stress along the *x*-axis induces drift motion along the z-axis, which helps bacteria approach the top and bottom surfaces. With our previous study [78] and theory by Mathijssen et. al [54], the bacterial reorientation rate on the surface in the presence of flow can be described as:

$$\boldsymbol{\Omega} = \boldsymbol{\Omega}^W + \boldsymbol{\Omega}^F + \boldsymbol{\Omega}^V + \sqrt{2D_r}\boldsymbol{\xi}^r,$$

The total rotational rate, $\boldsymbol{\Omega}$, is primarily contributed by the surface term $\boldsymbol{\Omega}^W$, the local shear flow term $\boldsymbol{\Omega}^F$, and the surface-flow coupled effects $\boldsymbol{\Omega}^V$. Specifically, $\boldsymbol{\Omega}^W$ arises from the hydrodynamic and steric interactions between the bacteria and the surface. The flow mainly affects the Jeffery orbit $\boldsymbol{\Omega}^J$ and the chirality-induced drift motion $\boldsymbol{\Omega}^C$, such that $\boldsymbol{\Omega}^F = \boldsymbol{\Omega}^J + \boldsymbol{\Omega}^C$. The surface-flow coupled effect, $\boldsymbol{\Omega}^V$, is primarily contributed by the weathervane effect, which also corresponds to bacterial upstream swimming.

$D_r$ includes both Brownian rotational diffusion and active rotational diffusion due to bacterial tumbling. $\boldsymbol{\xi}^r$ represents independent white noise stochastic processes with zero mean and correlation δ(t). Among these parameters, $\boldsymbol{\Omega}^W$, $\boldsymbol{\Omega}^C$, $\boldsymbol{\Omega}^V$ promote bacterial surface accumulation, while $\sqrt{2D_r}\boldsymbol{\xi}^r$ facilitates bacterial escape. The role of $\boldsymbol{\Omega}^J$ in trapping or escaping depends on bacterial shape and shear rate. Overall, bacterial surface accumulation is largely governed by physical factors that are highly responsive to bacterial morphology and flow conditions. The balance between these factors determines bacterial orientation on the surface and ultimately the total number of bacteria that accumulate.

**Conclusions and Perspective**

In summary, our research expands the experimental study of bacterial motion and accumulation from simplified surfaces to biotic surfaces, exploring both the similarities and differences. For surface-swimming bacteria, fluctuations in the cell surface reduce the effective height between the bacteria and the surface, resulting in a decreased radius of circular motion. Additionally, adherent bacteria are observed on cell surfaces, which can be categorized into two distinct types: tight adhesion and loose adhesion. These types exhibit significant differences in their occurrence positions, diffusion regions, and resistance to flow. Tight adherent bacteria predominantly cluster along cell edges and are resistant to flow-induced detachment, whereas loosely adherent bacteria are more likely to be displaced by the flow. Furthermore, classical surface accumulation effects were also explored. Our results show that cell surfaces accumulate more bacteria, primarily due to an increase in adhesion. Under flow conditions, accumulation is intensified, but this is driven by shear-induced hydrodynamic effects. Regardless of flow conditions, bacterial accumulation displays a time-dependent pattern, with flow conditions promoting greater accumulation by bringing fresh bacteria to the surface.

Overall, at the single-cell layer level, bacterial swimming and accumulation are qualitatively dominated by physical effects, although biological processes may also have some influence from a quantitative perspective. Furthermore, at tissue and organismal levels in vivo, cells not only establish tight junctions to perform critical biological functions, such as serving as barriers and facilitating endocrine activities, but also interact with various immune cells and participate in inflammatory responses within the complex microenvironment, the competition between biochemical and physical factors presents an intriguing area for further exploration. These questions are crucial for understanding bacterial colony formation and the onset of infections in biological environments.

**Methods and Materials**

**Fabrication of microfluidic channel.** Microchannels, each 600 microns in width, 100 microns in height, and 3 cm in length, were constructed by a 10:1 v/v ratio mixture of Polydimethylsiloxane (PDMS, Dow Corning) as shown in Fig. 1a. After treatment with a plasma cleaner, the channels were bonded to 24x50mm glass slides. One side of the microchannel a hole with 0.7mm in diameter was punctured to connect the syringe pump for introducing a uniform flow field and *E. coli* RP437 bacterial solution. On the other side of the microchannel, a 3-mm hole was punched to serve as a reservoir for medium during endothelial cell culture to provide nutrients for cell growth. Prior to cell seeding, all channels were modified with 25ng/mL fibronectin (FN) for 12 hours at 4 °C to promote endothelial cell adhesion.

**Cell culture.** Human umbilical vein endothelial cells (HUVECs) were purchased from American Type Culture Collection (ATCC). HUVEC-GFP were commercially obtained and cultured in endothelial cell medium (ECM, ScienCell). A seeding density of $5 \times 10^6$ endothelial cells/mL in the channel corresponded to 100% density of endothelial cells. Endothelial cells were inoculated and cultured in microchannels for 8 hours to ensure their adherence before introducing bacteria.

**Preparation of bacteria.** In this experiment, fluorescently labeled *E. coli* RP437 was used. Initially, 10 μL of bacteria preserved at -80 °C were extracted and cultured in Luria-Bertani (LB) broth. The medium was incubated overnight at 30 °C and 200 rpm until bacteria reached the mid-exponential phase. LB broth was consisting of 1.0% (w/v) Tryptone, 0.5% (w/v) Yeast Extract, and 1.0% (w/v) NaCl, dissolved in deionized water. Subsequently, the bacterial suspension incubated further in fresh LB broth at a 1:100 ratio until the bacterial concentration achieved an optical density (OD) of approximately 0.6 at a $\lambda = 600$ nm wavelength. To maintain bacterial motility while preventing division, the bacteria was then resuspended in a motility buffer composed of 1M sodium lactate, 100 mM EDTA, 1 mM L-methionine, and 0.1 M potassium phosphate buffer (pH 7.0).

**Microscopic imaging and data analysis.** All images in our experiment were captured using an inverted fluorescence microscope (Eclipse Ti2, Nikon). Images of bacteria were captured at 30 fps. Bacterial trajectories and speeds were analyzed using the Trackmate plugin in FIJI and further processed and mapped using MATLAB (the Mathworks).